\documentstyle[12pt]{article}
\begin{document}
%\setlength{\textheight}{20 cm}
%\begin{document}
\begin{titlepage}
\begin{flushright}
TP-USl/96/11
\vspace{1 cm}
\end{flushright}
\begin{center}
{\LARGE\bf Physics of heavy neutrinos$^{\ast}$} \\
\vspace{1.5 cm}
{\large\bf J. Gluza}\footnote{e-mail address: gluza@us.edu.pl} 
and {\large\bf M. Zra{\l}ek}\footnote{e-mail address: zralek@us.edu.pl} \\
\vspace{ 0.5 cm}
Department of Field Theory and Particle Physics \\
Institute of Physics, University of Silesia \\
Uniwersytecka 4, PL-40-007 Katowice, Poland \\
\vspace{1cm}
%\maketitle
\baselineskip 1 mm
{\large\bf Abstract} \\
\end{center}
Theoretical and experimental situation in physics of heavy 
neutrinos $(M_N>M_Z)$ is briefly presented. Various experimental
bounds on heavy neutrino masses and mixings are shortly reviewed.
Special attention is 
paid to possibility of detecting heavy neutrinos in future lepton 
\vspace{2 cm} linear colliders.

$^{\ast}$ Lecture presented by J. Gluza at the International Conference 
on Particle
Physics and Astrophysics in the Standard Model and Beyond, Bystra, September
1995.
\end{titlepage}
\vspace{0.5 cm}
\baselineskip 6 mm
\section{Introduction}

After the discovery of the top quark, neutrinos (and Higgs particle - not 
observed yet) remain the most elusive particles. Since very beginning 
neutrinos have played an important part in our understanding of the laws of
particle physics. They are the only particles which interact by only one type
of fundamental interaction, the weak one. The weak interaction of the other
particles is suppressed by their electromagnetic and strong ones.
To understand how important the weak interaction of neutrinos is in Nature (especially 
for us) let's mention just the mechanism
in which the Sun is shining. Without any doubt investigation of properties
of these particles can reveal many interesting, hidden until now physical
phenomena or explain many hypothetical ideas. For instance existence of
nonzero mass of neutrinos, besides theoretical interest, could explain some 
astrophysical or cosmological
problems [1]. This feature of neutrinos would boost particle physics beyond the 
Standard Model
(SM). Until now we know only that three known neutrinos are very
light ($m_{\nu_e} \leq$ few eV, $m_{\nu_{\mu}} \leq $ 270 keV, 
$m_{\nu_{\tau}} \leq$ 24 MeV).

Many models beyond the SM predict massive neutrinos. Moreover, except the
light neutrinos they predict very heavy ones: this is for instance in the
case of so-called `see-saw' models [2].

In this talk we give the short review of theoretical and experimental situation
connected with heavy neutrino physics. In the next Section we specify areas
where heavy neutrinos could reveal themselves. Section 3 will be devoted
specially to future linear $e^+e^-$ and $e^-e^-$ colliders and possibility of
finding heavy neutrinos there. These considerations will be given in the frame
of the simplest extension of the SM including right-handed neutrinos.

Conclusions and outlook will be given in the Summary.

\section{ Where to hunt for a very heavy neutrino?}
%\subsection{Cross section in the frame of the RHS model.}

Heavy neutrinos have been looked for since the early seventies [3]. From the
negative search of new neutral states and from the measurement of Z decay width
at the LEP we know that there are no neutrinos with a standard coupling to Z and mass
below $M_Z/2$ [4] or even below $M_Z$ if $BR(Z^0 \rightarrow \nu N) >
3 \cdot 10^{-5}$ [5]. The lack of detection of new neutrino states at the LEPI
indicates that if they exist, they will generally have large masses
($ \geq M_Z$).

Let's describe shortly where such heavy neutrinos could be found. 

\subsection{Influence of heavy neutrino states on observables measured at LEP
and low energy experiments}

Even if they can not be directly produced at LEP now, it is still 
possible that with increasing precision of measurements 
their effects could be indirectly detected as small deviations of couplings
of light neutrinos from their standard values. This could happen for any
new neutrino states if they mix with the ordinary ones.
As an example let us mention the observation made 
by C. Jarlskog [6] and discussed by other authors [7].

~In the SM with n left-handed lepton doublets and $n'=1,2,...$ right-handed 
neutrinos the effective number of neutrino species $\eta_{exp}$ measured at the
$Z^0$ peak and defined by ($\Gamma_0$ is the SM $Z^0$ decay width to the pair of
massless neutrinos)
$$\Gamma ( Z^0 \rightarrow neutrinos ) = \Gamma_0 \eta_{exp}$$
fullfils relation
$$\eta_{exp} \leq n.$$

That means that any measured value $\eta_{exp}$ slightly below 3 (number of 
left-handed lepton doublets in the SM) would indicate existence of right-handed
neutrinos (at the moment the best fit for $\eta_{exp}$ is $\eta_{exp}=2.991 \pm 0.016$ [4]).

Similar phenomena could be observed for other LEP observables as $\Gamma_Z$,
Z partial decay widths, asymetries measured at the Z resonance, W-mass and
low energy experiments as $\beta,
\;\tau$ and $\pi$ decays, $\nu-$scattering, atomic parity violation,
polarized e-D scattering, etc. Global analysis of fermion mixings
with new neutral states can be found in [3,8].

\subsection{Heavy neutrinos in hadronic colliders}

This possibility for detecting heavy neutrinos was examined in many papers [9-13].
The especially big hope was connected with construction of the Superconducting
Super Collider (SSC) [10]. After cancellation of this project 
the possible options which remain are $e^-p$ and $pp$ at HERA, HERA upgrade
and LEP+LHC colliders. According to 
[11] masses up to $\sim$ 160 GeV, 320 GeV and 700 GeV can be tested in ep 
collisions at HERA, HERA upgrade and LEP+LHC, respectively.
Let's note however that such 
optimistic results for $e^-p$ collider are predicted for very large mixing 
angle $\xi=0.1$
which is much above up-to-date constraints on $\xi$ (see the next section).
$pp$ super-colliders could give detectable results through $pp \rightarrow W_R 
\rightarrow l^+N_l \rightarrow l^+l^+q\bar{q'}$ reaction (quark fusion) [12] or
through the gluon fusion mechanism with off-shell Z gauge boson 
($gg \rightarrow  Z^{\ast} \rightarrow N\bar{N}$) [13].

\subsection{Induced heavy neutrino loop effects}

Some authors [14] indicate that significant rates are in general possible for 
one-loop-induced rare processes as $\mu-e$ conversion in nuclei, $\mu (\tau)
\rightarrow 3e$, $\mu \rightarrow e \gamma$ due to exchange of virtual heavy neutrinos. 
The possibility of detecting such lepton number violating processes could arise
when heavy neutrinos do not decouple in low energy processes. This can happen
in other than `see-saw' models [3,15].

\subsection{Neutrinoless double-beta decay}

The search for neutrinoless double-$\beta$ decay $\left( (\beta\beta)_{0\nu}
\right)$
$$(A,Z) \rightarrow (A,Z \pm 2) + 2e^{\mp}$$
is the most promising method for the discovery of light Majorana masses.
The reaction is also sensitive to heavy neutrinos' 
contribution. 
There are about 40 different experiments being carried out now in which 
people are looking for this type of reaction. 
%The reaction is also sensitive to heavy neutrinos' 
%neutrinos can become apparent only through detailed comparison of decays
%involving various nuclei what seems to be practically impossible. 
%As this process hasn't been observed yet, only constraints on heavy neutrino
%mixings and masses are derived (see the next section).
 
\subsection{Indirect detection of heavy neutrinos in neutrino oscillation 
experiments}

This interesting possibility was discussed in [16].
It was shown that if there is a see-saw type mixing between light and heavy 
Majorana
particles and the mixing matrix is complex then the $\nu_{\mu} \rightarrow
\nu_{\tau}$ and $\bar{\nu}_{\mu} \rightarrow \bar{\nu}_{\tau}$ transition 
probabilities could be different and indicate (indirectly) presence of heavy 
neutrinos. \\

None of the processes including heavy neutrinos described above have been
discovered till now and only some constraints on 
allowed heavy neutrino masses and mixings can be derived from them. The
most important constraints will be presented in the next section when we shall deal
in details with the last, very important area where heavy neutrinos can be 
found - future linear lepton colliders. To complete the review let's mention
that a heavy neutrino is naturally highly unstable so no discrepancy with
cosmology appears here.

\section{Heavy neutrinos in the Next Linear Colliders}

Recently hadron colliders gave spectacular results when W and Z bosons and
the top quark were discovered. Nevertheless, in the meantime a lepton collider,
LEPI, has reached spectacular results too, specially these connected with 
excellent
precision with which the SM has been tested. The next planned $e^+e^-$ colliders
with energy up to 2 TeV [17] can become even more important as a tool in looking 
for new physics beyond the SM, for instance connected with detection of heavy
neutrinos. The last part of this talk will be devoted to the physics of heavy
neutrinos in these future colliders.

We'll focus on two reactions: the direct heavy neutrino $e^+e^- \rightarrow
\nu N$ production and the indirect process with heavy neutrino exchange
$e^-e^- \rightarrow W^-W^-$. This latter process is possible as the $e^-e^-$
option of the next linear accelarators and is seriously taken into account [18].
The $e^-e^-$ environment is much cleaner than the $e^+e^-$ one. There is
much less SM activity and that is why it allows to explore even very weak
signals of flavour violating processes as this given above. The values of cross
sections which we are going to find depends on the model in which
we calculate them. So called `see-saw' models belong to the most popular ones
as they can give an theoretical explanation for a smallness of known neutrino 
masses [2].

As an illustration let's take the simplest model with massive neutrinos - the
SM with additional right-handed neutrinos (RHS model). 
In the RHS model there are 3 left-handed and $n_R$
(=1,2,...) right-handed weak
neutrino states transforming under $SU_L(2)$ gauge group
as doublets and singlets,
respectively. The neutrino mass matrix has $3+n_R$
dimensions
\begin{equation}
M_{\nu}= { \overbrace{0}^{3} \ \overbrace{M_D}^{n_R} \choose M_D^T \ M_R }
{\begin{array}{c}
\} 3 \\ \} n_R.
\end{array}}
\end{equation}
Without Higgs triplet fields the $3 \times 3$ dimension part $M_L$ of
$M_{\nu}$ equals zero
\begin{equation}
M_L=0.
\end{equation}
Using ($3+n_R$) dimensional unitary matrix 
\begin{equation}
U=\left( \matrix{ K^T \cr
               U_R } \right) 
\end{equation}                                                      
which acts on the weak neutrino states, we can diagonalize matrix $M_{\nu}$
($U^TM_{\nu}U=M_{diag}$) and get the
physical states. 

Without loosing the
generality we can assume that the charged lepton mass matrix is diagonal, so
then the physical neutrino $N= \left( N_1,...,N_{3+R} \right)^T$
couplings to gauge bosons are defined by
($\hat{l}=(e,\mu,\;\tau)^T,\;P_L=\frac{1}{2}(1-\gamma_5)$)
\begin{eqnarray}
L_{CC}&=&\frac{g}{\sqrt{2}}\bar{N}\gamma^{\mu}KP_L\hat{l}W_{\mu}^+ + h.c., \\
L_{NC}&=&\frac{g}{2\cos{\theta_W}} \left[
\bar{N}\gamma^{\mu}P_L(KK^{\dagger})N \right].
\end{eqnarray}
For instance for $n_R=3$ we get three light (known) neutrinos and three very
heavy $M_{1,2,3} \geq M_Z$ ones as $M_R$ and $M_D$ are proportional to
different scales of symmetry breaking and 
$\mid M_R \mid_{ii} >> \mid M_D \mid_{lk}.$ Then without any additional symmetry
the matrix elements $K_{ae}$ are proportional to $<M_D>/M_a$. Typically 
$<M_D> \sim$ 1 GeV so $K_{ae}$ is very small and very sensitive to the $M_a$ 
mass. The process $e^-e^+ \rightarrow \nu N$ is proportional to $\mid K_{ae} 
\mid^2$ [19] and the $e^-e^- \rightarrow W^-W^-$ to $\mid K_{ae} \mid^4$ [20] and
typical cross sections as a function of $M_N$ for different $\sqrt{s}$ energies
are given in Fig.1 (taken from [19]) and Fig.2. One can see that it is not 
possible to detect the $e^-e^- 
\rightarrow W^-W^-$
process (the `detection limit' on the $\sigma$=0.1 fb level 
is reasonable for this process [21]).
The cross section for the $e^+e^- \rightarrow \nu N$ process is small.
However, the `see-saw' mechanism is not the only scenario which explains
small masses of the known neutrinos. There are models based on symmetry
argument [15] where no simple relations connected $M_a$ with $K_{ae}$ 
are present. In this case the mixing matrix elements are independent 
parameters and as such are bounded only by experimental data. There are
four different and important sources of constraints on heavy neutrino mixings
coming from experiments.

(i) From LEPI we know that 
if neutrinos with masses below $M_Z$ exist their couplings
to $Z_0$ should be such that $Br( Z \rightarrow N \nu) \leq 3 \cdot 10^{-5}$ [5]
(what is equivalent to $K_{ae}^2 \leq 8\cdot 10^{-5}$). 
Because this mixing is very small we resign from study very tiny effects 
connected with neutrinos with $M_N < M_Z$ and we'll only study the case $M_N 
\geq M_Z$.

(ii) Low energy experiments (e.g. lepton universality, the $\mu$ decay)
and LEPI give also information about heavy neutrinos with masses above $M_Z$.
The reason is that due to unitarity properties of the U matrix (Eq.(3)), 
the nonzero mixing 
matrix elements $K_{ae}$ slightly reduce the couplings of light neutrinos from 
their SM 
values thus affecting all processes including neutrinos [3] (in the SM matrix K
in Eqs.(4,5) equals I). The up-to-date
limit for RHS model is [22] 
\begin{equation}
\kappa^2=\sum\limits_{a=heavy}K_{ae}^2 \leq 0.0054.
\end{equation}

(iii) The lack of signal of neutrinoless double-$\beta$ decay $(\beta\beta)_{0\nu}$ gives
the
bound for light neutrinos
\begin{equation}
\left| \sum_{\nu(light)}K_{\nu e}^2m_{\nu} \right| <
\kappa^2_{light}
\end{equation}
where $\kappa^2_{light}<0.68$ eV [23]. \\

(iv) From the $\left( \beta\beta \right)_{0\nu}$ process it is also possible
to get the bound for heavy neutrinos 
\begin{equation}
\left| \sum\limits_{N(heavy)}K_{Ne}^2\frac{1}{M_N} \right| < \omega^2.
\end{equation}
Typically the bound is: $\omega^2<5.6\cdot10^{-4}\rm\;TeV^{-1}$~[24]. \\

The last constraint which we use comes from the fact that the mass term for
the left-handed neutrinos is absent

(v) 
\begin{equation}
\sum\limits_{\nu (light)}K_{\nu e}^2m_{\nu}+\sum\limits_{N(heavy)}K_{Ne}^2M_N 
=M_L \equiv 0.
\end{equation}

This fact confronted with Eq.(7) gives
\begin{equation}
\left| \sum_{N(heavy)}K_{Ne}^2M_N \right| < \kappa^2_{light}.
\end{equation}
This relation includes an interesting information. To get meaningful values of 
cross sections for the studied processes we need
the values of $K_{Ne}$ as big as possible. As $\kappa^2_{light}$ in Eq.(10) is 
very small
the only possibility to reconcile these two facts is to assume that some 
$K_{Ne}$ matrix elements are complex numbers. 
If CP symmetry is conserved then complex $K_{Ne}$ numbers are equivalent 
to the fact that $\eta_{CP}$ parities of heavy neutrinos are not all equal.

Now we deduce that if CP parities of all heavy neutrinos are the same or we have only one
right-handed neutrino ($n_R=1$) then both considered processes are very small.
Situation is different if $n_R=2$. In agreement with our discussion let's take
heavy neutrinos with opposite CP parities $\eta_{CP}(N_1)=-\eta_{CP}(N_2)=i$
and masses $M_1=M,\;M_2=AM\;(A \geq 1)$. Then taking into account Eqs.(6)-(10) 
the biggest mixing angle $K_{N_1e}$ is for $A \rightarrow 1$ (for details see 
[20]). The result is shown in Fig.3 (taken from [20]) for the $e^+e^- \rightarrow \nu N$ process.
%3
The solid line represents the biggest result and does not change for $n_R >2$.
\footnote{The biggest possible $K_{Ne}$ is [20,25] ${(K_{Ne})}_{max} \simeq
\frac{\kappa^2}{2}=0.0027$, that is why $\xi \equiv K_{Ne}=0.1$ as mention
in Section 2.2 is too big.}
However the $e^-e^- \rightarrow W^-W^-$ process still remains below the 
detection limit.
This is because for $A \rightarrow 1$ we have two degenerate
Majorana neutrinos $(M_1=M_2)$ with opposite CP parities which is equivalent 
to one Dirac 
neutrino.

The case with $n_R=3$ changes situation for the $e^-e^- \rightarrow W^-W^-$ 
process. In Fig.4 (taken from [20]) we show the most optimistic results for the $e^-e^- 
\rightarrow W^-W^-$ cross section. Taking $\eta_{CP}(N_1)=
\eta_{CP}(N_2)=-\eta_{CP}(N_3)=i$ and $M_1=M,\;M_2=AM,\;M_3=BM$ we found
values A,B for which $\sigma( e^-e^- \rightarrow W^-W^-)
$ reaches maximum. This situation takes place for the very heavy second
$(A >>1)$ and heavier third neutrino ($B \sim 2-10$). In this Figure we
depict also the cross section for production of the lightest heavy neutrinos
with the mass M in the $e^+e^- \rightarrow \nu N$ process taking exactly the same
mixing angle $K_{N_1e}$ as for the $e^-e^- \rightarrow W^-W^-$ process. \\
%4

We can conclude that 

(i) everywhere in the possible region of phase space the
production of heavy neutrinos in the $e^+e^-$ process has 
greater cross section than the lepton violating process
$e^-e^-$. It is impossible to find such mixing angles and masses that 
would show the opposite.
The large values of $\sigma \left( e^+e^-
\rightarrow N\nu \right) $ make this process a good place for
the heavy neutrino searching and for future detailed
studies (decay of heavy neutrinos, background from other
channels [25]). 

(ii) there are also regions of heavy neutrino masses outside
the phase space region for $e^+e^-$ where the $\Delta L=2$ process
$e^-e^-$ is still a possible place to look for heavy neutrinos. It
is a small region $1\;{\rm TeV}<M<1.1\;{\rm TeV}$ for $\sqrt{s}=1$~TeV,
$1.5{\rm\;TeV}<M<2{\rm\;TeV}$ for $\sqrt{s}=1.5$ TeV and
$2{\rm\;TeV}<M<3.1{\rm\;TeV}$ for $\sqrt{s}=2$~TeV where the cross section
$\sigma \left( e^-e^- \right) $ is still above the `detection
limit'. There is no such place with the $\sqrt{s}=0.5$ TeV
collider. The experimental value of $\kappa^2$ (see Eq.(6)) would have to
be below $\sim 0.004,\sim 0.003,\sim 0.002$ for
$\sqrt{s}=1,1.5,2$ TeV respectively to cause these regions
to vanish. 

The largest value of the mixing parameter $\mid K_{Ne} \mid$ for
$n_R>3$ is the same as in the $n_R=3$ case and we do not obtain 
quantitatively new results in these cases.

To sum up, we have found the `maximum possible' cross sections for
production of the heavy
neutrino ($e^+e^- \rightarrow N\nu$ process) and for the inverse neutrinoless
double-$\beta$ decay
($e^-e^- \rightarrow W^-W^-$ process) in the energy range interesting for
future lepton colliders
(0.5--2 TeV). The upper values for the cross sections are still large enough
to be
interesting from an experimental point of view. For the $e^+e^- \rightarrow
N\nu$
process the cross
section could be as large as 275 fb for $\sqrt{s}=1$ TeV and
$M=100$ GeV. The $e^-e^-
\rightarrow W^-W^-$ process could give indirect indication for larger massive
Majorana neutrino
existence, not produced in the $e^+e^- $ scattering. 

\section{Summary}

In this talk we review the possibilities of detecting heavy neutrinos which are
present in plenty of theoretical models beyond the SM. None of the nonstandard
processes involving heavy neutrinos has ever been detected. However, on
theoretical ground, narrow windows are still open even after taking into account
up-to-date stringent limits on heavy neutrino mixing angles and masses. The
most promising are reactions with the ep hadron colliders and the 
$e^+e^-$ accelarators.
Indirect signals of heavy neutrinos presence can be looked for in induced 
by them loop processes as $\mu \rightarrow e \gamma$, $\mu (\tau) \rightarrow 3e$ 
and in the future $e^-e^-$ accelarators.

\section*{Acknowledgments}

Supported by the Polish Committee for Scientific Research under the grant No 
PB659/P03/95/08.
%%%%%%%%%%%%%%%%%%%%%%%%%%%%%%%%%%%%%%%%%%%%%%%%%%%%%%%%%%%%%%%%%%%%%%%%%%%%
\section*{References}
\newcounter{bban}
\begin{list}
{$[{\ \arabic {bban}\ }]$}{\usecounter{bban}\setlength{\rightmargin}{
\leftmargin}}
\item G.Gelmini, E.Roulet, Rep.Prog.Phys.58(1995)1207.
\item T. Yanagida, Prog.Theor.Phys.{\bf B135} (1978) 66; M.~Gell-Mann,
P.~Ramond and R.~Slansky, in `Supergravity', eds. P.~Nieuwenhuizen and
D.~Freedman (North-Holland, Amsterdam, 1979) p.315.
\item H.B.Thacker and J.J.Sakurai, Phys.Lett.{\bf B36}(1971)103,
Y.S.Tsai, Phys.Rev.{\bf D4}(1971)2821; J.D.Bjorken and C.H.Llewellyn
Smith, Phys. Rev. {\bf D7}(1973)887, 
M.Gronau, C.N.Leung, J.L.Rosner Phys.Rev.{\bf D29}(1984)2539;
P.Langacker, D.London Phys.Rev.{\bf D38}(1988)886.
\item The LEP Collaborations ALEPH, DELPHI, L3, OPAL and the LEP Electroweak
Working Group, CERN-PPE/95-172.
\item L3 Collaboration, O. Adriani et al., Phys. Lett. {\bf B295} (1992) 371
and {\bf B316} (1993) 427.
\item C.Jarlskog Phys.Lett.{\bf B241}(1990)579.
\item S.M.Bilenky, W.Grimus, H.Neufeld Phys.Lett.{\bf B252}(1990)119,
\newline C.O.Escobar et al. Phys.Rev.{\bf D47}(1993)R1747.
\item E.Nardi, E.Roulet, D.Tommasini Nucl.Phys.{\bf B386}(1992)239, 
\newline Phys.Lett.{\bf B344}
(1995)225, C.P.Burgess et.al Phys.Rev.{\bf D49}(1994)6115.
\item W.Keung, G.Senjanovic Phys.Rev.Lett.50(1983)1427.
\item D.A.Dicus, D.D.Karatas, P.Roy Phys.Rev.{\bf D44}(1991)2033,\newline
B.Mukhopadhyaya Phys.Rev.{\bf D49}(1994)1350, H.Tso-hsiu, Ch.Cheng-rui, T.Zhi-jian
Phys.Rev.{\bf D42}(1990)2265.
\item W.Buchmuller, C.Greub Nucl.Phys.{\bf B363}(1991)345, {\bf B381}(1992)109,
G.Ingelman, J.Rathsman Z.Phys.{\bf C60}(1993)243,
A.Djouadi, J.Ng, T.G.Rizzo hep-ph/9504210.
\item A.Datta, M.Guchait, D.P.Roy Phys.Rev.{\bf D47}(1993)961.
\item D.A.Dicus, P.Roy Phys.Rev.{\bf D44}(1991)1593.
\item T.P.Cheng and L.F.Li,Phys.Rev.{\bf D44}(1991)1502,
A.Ilakovac and
A.Pilaftsis Nucl.Phys.{\bf B437}(1995)491,
D.~Tommasini, G.~Barenboim, J.~Bernabeu and
C.~Jarlskog, Nucl. Phys. {\bf B444} (1995) 451.
\item D. Wyler and L. Wolfenstein, Nucl. Phys. {\bf B218} (1983) 205;
R.N.~Mohapatra and J.W.F.~Valle, Phys. Rev. {\bf D34} (1986) 1642;
E.~Witten, Nucl. Phys. {\bf B268} (1986) 79;
J.~Bernabeu et al., Phys. Lett. {\bf B187} (1987) 303;
J.L.~Hewett and T.G.~Rizzo, Phys. Rep. {\bf 183} (1989) 193;
E.~Nardi, Phys. Rev. {\bf D48} (1993) 3277.
\item S.M.Bilenky, C.Giunti Phys.Lett.{\bf B300}(1993)137.
\item See e.g. R.Settles, `$e^+e^-$ Collisions at 500 GeV: The Physics
Potential' edited by P.M. Zerwas, DESY 93-123C.
\item See e.g. Proc.of the Workshop on physics and experiments with linear
colliders (Saariselk$\ddot{a}$, Finland, September 1991),edited by
R.Orava,P.Eerola and
M.Nordberg (World Scientific, 1992) and Proc.of the Workshop on physics
and experiments with linear colliders (Waikoloa,Hawaii,April 1993),edited by
F.A.Harris, S.L.Olsen, S.Pakvasa and X.Tata (World Scientific, 1993).
\item J. Gluza and M. Zra{\l}ek Phys.Lett.{\bf B362}(1995)148.
\item J.Gluza and M. Zra{\l}ek Phys.Lett. {\bf B372}(1996)259.
\item J.F.Gunion and A.Tofighi-Niaki, Phys.Rev.{\bf D36},2671(1987) and
{\bf D38},1433(1988), F.Cuypers, K.Ko{\l}odziej, O.Korakianitis and
R.R$\ddot{u}$kl Phys.Lett.{\bf B325}(1994)243.
\item A.Djouadi et. al. in [11].
\item A. Balysh et al. (Heidelberg-Moscow Coll.), Proc. of the International
Conference on High Energy
Physics, 20--27 July 1994, Glasgow, ed. by P.J.~Bussey and I.G.~Knowles,
vol.II, p.939.
\item J.D. Vergados, Phys. Rep. {\bf 133} (1986) 1.
\item J. Gluza, D. Zeppenfeld, M.Zra\l ek, in preparation.
\end{list}
%\newpage
%\end{document}
%%%%%%%%%%%%%%%%%%%%%%%%%%%%%%%%%%%%%%%%%%%%%%%%%%%%%%%%%%%%%%%%%%%%%%%%%%%%%%
\section*{Figure Captions}
\newcounter{bean}
\begin{list}
{\bf Fig.\arabic
{bean}}{\usecounter{bean}\setlength{\rightmargin}{\leftmargin}}
\item The cross section for the $e^-e^- \rightarrow W^-W^-$ process as a
function of the heavy neutrino mass for the `classical'
see-saw models, where the mixing angles between light and heavy neutrinos are
proportional to the inverse of mass of the heavy neutrino. 
Solid (dashed) line is for the TLC (NLC) collider's energy.
\item  The cross section for the $e^-e^+ \rightarrow \nu N$ process in the 
frame of the `classical'
see-saw models. Solid, dashed lines and line with stars are for 
1 TeV, 500 GeV and 200 GeV CM energies of future colliders respectively.
\item The cross section for the $e^+e^- \rightarrow N\nu$ process as a function
of heavy
neutrino mass $M_1=M$ for $\sqrt{s}=1$ TeV in the models with two heavy
neutrinos ($n_R=2$) for
different values of $A=\frac{M_2}{M_1}$ (solid line with $A=1.0001$,
`$\diamond$' line with
$A=1.004$, dots line with A=1.01 and `$\ast$' line with A=100). Only for very
small mass difference $A
\sim 1$ existing experimental data leave the chance that the cross section
is large, e.g.\
$\sigma_{max}(M=100\;GeV) =275$~fb. If $M_2\gg M_1$ then the cross section must
be small, e.g.\ for
$A=100$, $\sigma_{max}(M=100 \rm\;GeV) \simeq0.5$ fb. The solid line gives also
$\sigma_{max}(e^+e^-\rightarrow N \nu)$ for $n_R>2$ (see the text).
\item The cross sections for the $e^+e^- \rightarrow N\nu$ and $e^-e^-
\rightarrow W^-W^-$
processes as a function of the lightest neutrino mass $M_1=M$ for different CM
energy (the
curves denoted by F05, F10, F15 and F20 depicted the cross section for both
processes
for $\sqrt{s}=$0.5, 1, 1.5 and 2 TeV respectively) for $n_R=3$.
The cross sections for the $e^-e^- \rightarrow W^-W^-$ process are chosen to be
the largest. For the
$e^+e^- \rightarrow N\nu$ reaction the cross section for each of neutrino
masses is calculated using
the same parameters as for $\sigma (e^-e^- \rightarrow W^-W^-)$ and is not the
iggest one (see
the solid line in Fig.~3 for the maximum of $e^+e^- \rightarrow
N\nu$). The solid line parallel to the
$M$  axis  gives the predicted `detection limit' $(\sigma=0.1\;$fb) for both
processes.
\end{list}

\newpage
\begin{figure}[p]
\vspace{6 cm}
\includegraphics{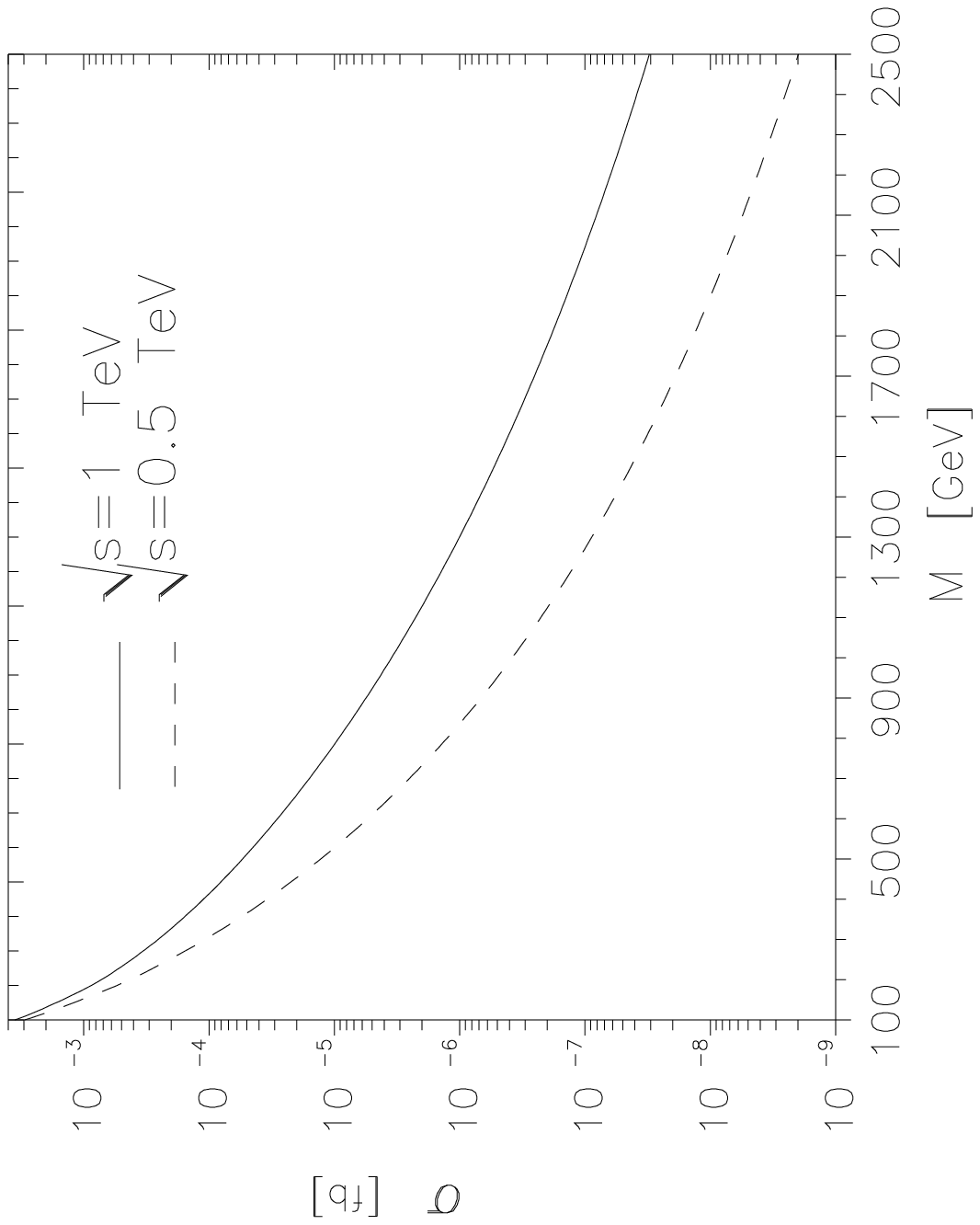}
\vspace{0.5 cm}
\caption{}
%\label{fig}
\end{figure}

\newpage
\begin{figure}[p]
\vspace{6 cm}
\includegraphics{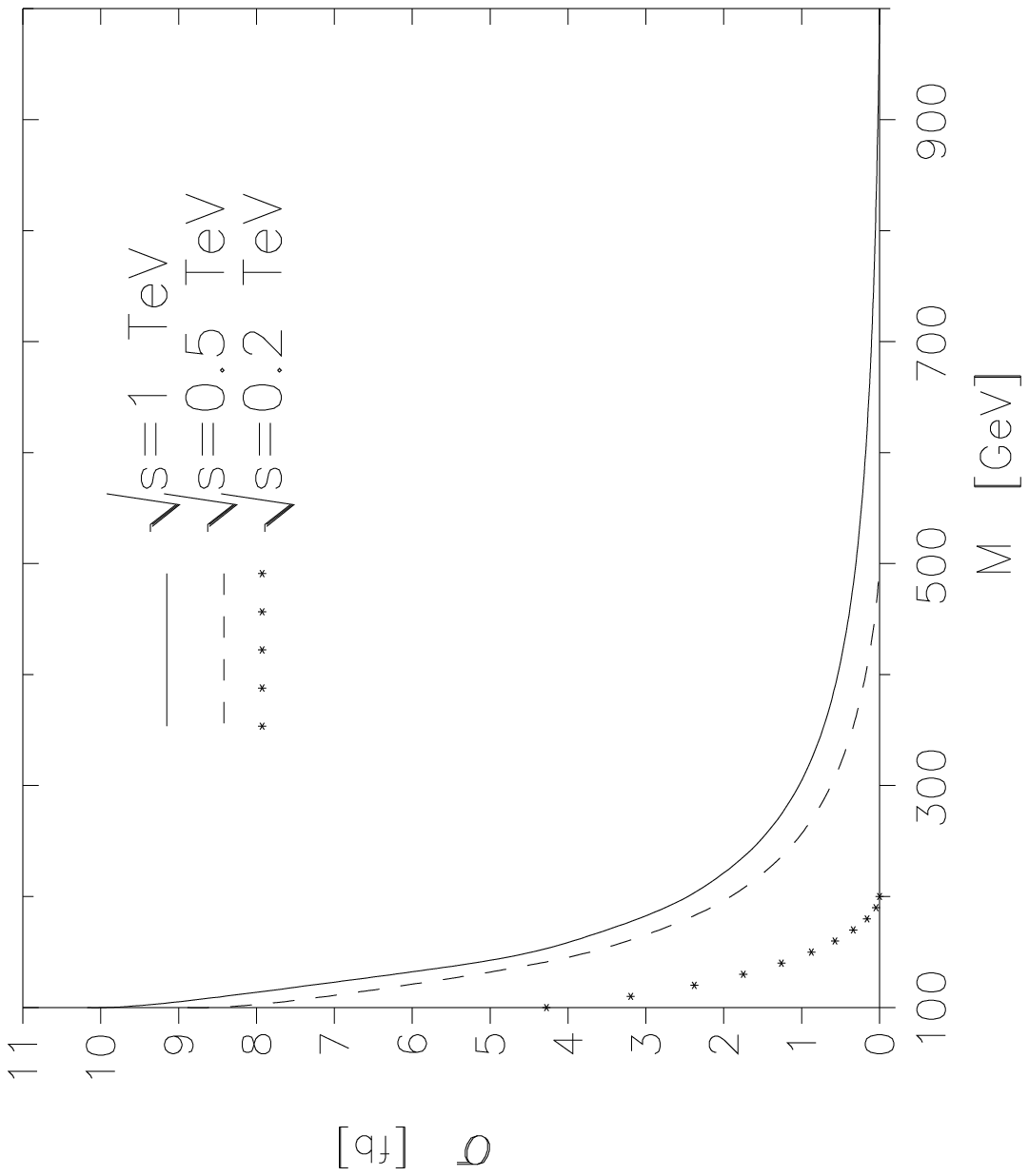}
\vspace{0.5 cm}
\caption{}
\end{figure}
\newpage
\begin{figure}[p]
\vspace{6 cm}
\includegraphics{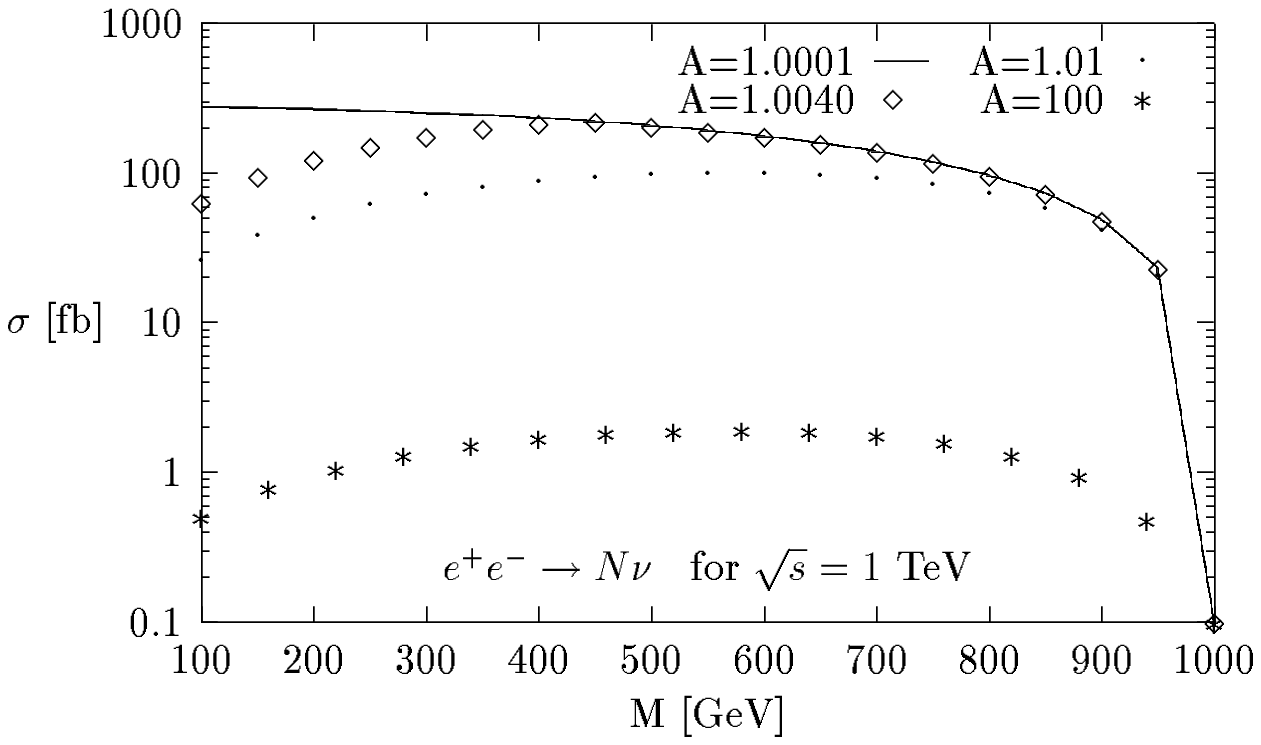}
\vspace{0.5 cm}
\caption{}
\label{fig:fig3}
\end{figure}
\newpage
\begin{figure}[p]
\vspace{6 cm}
\includegraphics{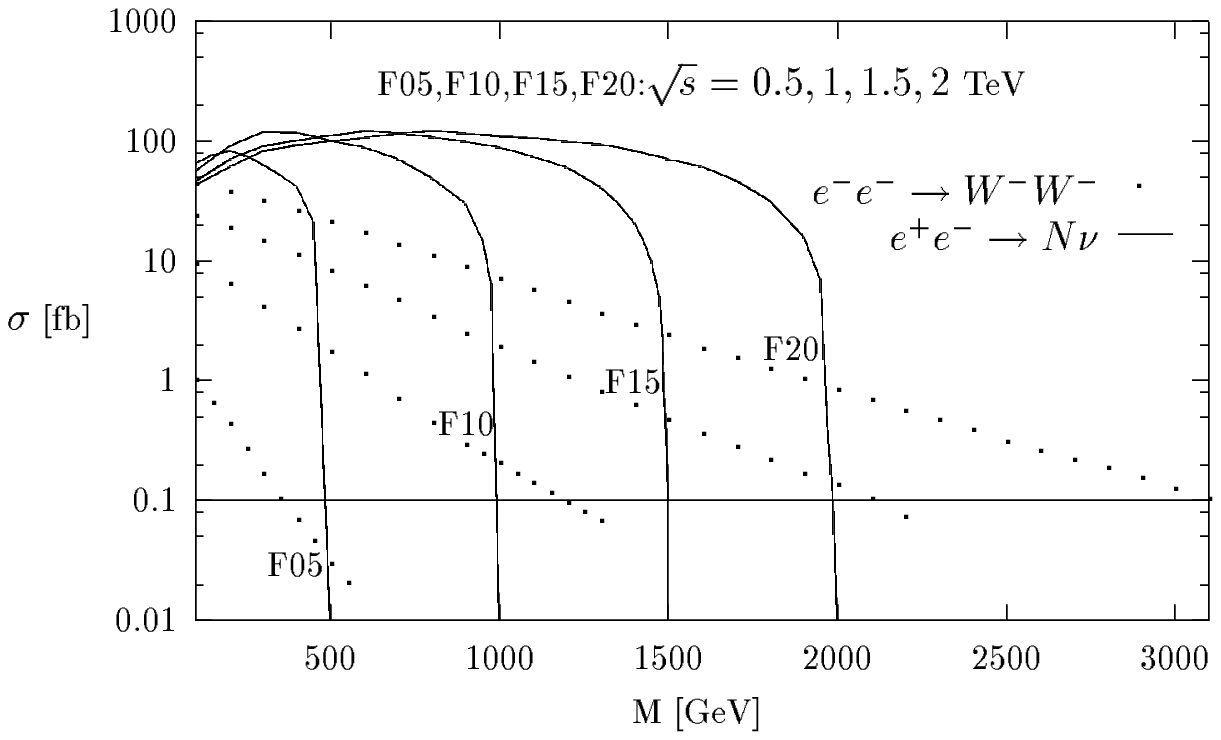}
\vspace{0.5 cm}
\caption{}
\label{fig:fig4}
\end{figure}
\newpage 
\end{document}